\documentclass[doublecol]{epl2} 
\usepackage{microtype}
\usepackage{amssymb}
\usepackage{graphics}

\newcommand{\be}{\begin{equation}}
\newcommand{\ee}{\end{equation}}
\newcommand{\om}{\omega}

\newcommand{\ra}{\rightarrow}
\newcommand{\D}{\mathrm{d}}

\newcommand{\cU}{\mathcal{U}}
\newcommand{\reals}{\mathbb{R}}
\newcommand{\rh}{\tilde h}
\newcommand{\rs}{\tilde s}
\newcommand{\cE}{\mathcal{E}}

\title{Ensemble equivalence for general many-body systems}
\shorttitle{Ensemble equivalence for general many-body systems}

\author{Hugo Touchette}
\shortauthor{H. Touchette}
\institute{School of Mathematical Sciences, Queen Mary University of London, London E1 4NS, UK}


\abstract{
It has been proved for a class of mean-field and long-range systems that the concavity of the thermodynamic entropy determines whether the microcanonical and canonical ensembles are equivalent at the level of their equilibrium states, i.e., whether they give rise to the same equilibrium states. Here we show that this correspondence is actually a general result of statistical mechanics: it holds for any many-body system for which equilibrium states can be defined and in principle calculated. The same correspondence applies for other dual statistical ensembles, such as the canonical and grand-canonical ensembles.
}

\begin{document}

\maketitle

\section{Introduction}

The microcanonical ensemble (ME) and canonical ensemble (CE) provide two different statistical descriptions of many-body systems at equilibrium -- the first in terms of their energy, the second in terms of the temperature of a heat bath connected to them. Although the two ensembles often give compatible or equivalent predictions in the thermodynamic limit, that equivalence is not guaranteed by their definitions. Gibbs noticed this point as early as 1902 \cite{gibbs1902} and, since then, many works have shown that the two ensembles either give equivalent or nonequivalent thermodynamic-limit predictions depending on the many-body system considered  or, more precisely, on the type of interactions acting on the particles or constituents of that system (see \cite{campa2009} for a comprehensive survey). 

The fundamental result that has emerged from these works is that many-body systems involving \emph{short-range} and stable interactions, such as short-range spin systems or screened Coulomb systems, always have equivalent ensembles, whereas systems involving \emph{long-range} interactions, such as gravitating systems, dipolar or unscreened Coulomb systems, may have nonequivalent ensembles \cite{campa2009}. How this equivalence or nonequivalence manifests itself depends on the level of description considered. At the \emph{thermodynamic} level, the ME and CE become nonequivalent when the heat capacity is negative in the ME but is positive in the CE. This surprising phenomenon, which has been studied first in the context of gravitational systems \cite{lynden1968,lynden1977,lynden1999,hertel1971,gross2001,ispolatov2001,chavanis2006}, is now known to arise when the thermodynamic entropy of the ME is nonconcave as a function of the mean energy. More recently, it has been shown that the ME and CE can also be nonequivalent if they have different sets of equilibrium states -- more precisely, if there exist equilibrium states in the ME, calculated at fixed energy, that do not arise in the CE at fixed temperature \cite{ellis2000}. When this is the case, we say that the ME and CE are nonequivalent at the \emph{macrostate} level. Various models have been shown to display this level of nonequivalence, including spin models \cite{barre2001,ellis2004,costeniuc2005a} and models of 2D turbulence \cite{ellis2002,venaille2009}.

Given that thermodynamics is ultimately rooted in statistical mechanics, it is natural to seek for a link between the thermodynamic and macrostate levels of ensemble equivalence, i.e., to determine whether the concavity of the entropy determines the macrostate equivalence of ensembles. This problem was studied by Eyink and Spohn \cite{eyink1993} and later by Ellis, Haven and Turkington \cite{ellis2000}, who proved that the ME and CE are, as physically expected, thermodynamically equivalent when they are equivalent at the macrostate level \cite{touchette2004b}. However, they were able to prove this result only for a limited class of systems that includes mostly mean-field-type systems. 

The aim of this paper is to show that the connection between the thermodynamic and macrostate levels of ensemble equivalence is actually a general result of statistical mechanics: \emph{it holds for any many-body system for which equilibrium states can be defined and in principle calculated}. This opens up the study of ensemble equivalence to many systems that do not fall in the class considered in \cite{ellis2000} -- most critically, gravitating systems, which are the source of much of the current research on long-range interactions \cite{chavanis2006,campa2009}. This general result also implies that the ME and CE are always equivalent for any macrostates of any short-range systems having concave entropies and, more generally, any systems that do not have first-order phase transitions in the CE.

The rigorous proofs of the results stated in this paper will be published elsewhere \cite{note1}. Here we focus on presenting the physical meaning of these results, as well as the main ideas involved in proving them, following \cite{ellis2000}. We illustrate the more general nature of our results by discussing the equivalence of the ME and CE for a simple spin model not completely covered by the results of \cite{ellis2000}. We also derive, from our equivalence results, general results about phase coexistence and first-order phase transitions. Finally, we discuss their generalization to ensembles other than the ME and CE.

\section{Framework}

We consider a general classical $N$-particle system described by the Hamiltonian or energy function $H(\om)$, where $\om\in\Lambda$ denotes the microstate of that system and $\Lambda$ the microstate space. The mean energy or energy per particle is denoted by $h(\om)=H(\om)/N$. We also consider a macrostate of that system, defined mathematically as a function $M(\om)$ of the microstates. This macrostate can represent, for example, the magnetization of a spin system, the empirical distribution of velocities or positions of a gas of $N$ particles, or a combination of any observables defined similarly as functions of $\om$.

The problem that we are concerned with is to define the set of equilibrium values of $M$ in both the ME and CE in the thermodynamic limit, and to determine whether these sets are equivalent, i.e., whether they can be put in some correspondence. Mathematically, the microcanonical equilibrium values of $M$, which we denote collectively as $\cE^u$, are identified as the typical values (viz., global maxima or concentration points) of the ME probability distribution $P^u(M=m)$ obtained by conditioning a uniform (prior) probability distribution $P$ on $\Lambda$ on the ``constrained'' set of microstates such that $h(\om)=u$. In symbols, this is expressed as
\be
P^u(m)=P(M=m|h=u)=\frac{P(M=m,h=u)}{P(h=u)}.
\ee
Note that, for simplicity, we use the same $P$ to denote the prior distribution $P(\om)$ on $\Lambda$ and probabilities in general. Thus $P(M=m)$ denotes the probability (density) of the event $M=m$ calculated with respect to $P(\om)$.

In a similar way, the canonical equilibrium values of $M$, denoted by $\cE_\beta$, are identified as the typical values of the CE probability distribution $P_{\beta}(M=m)$ of $M$ obtained by integrating the Gibbs distribution over the set of microstates such that $M(\om)=m$. Thus,
\be
P_{\beta}(M=m)=\int_{\om\in\Lambda:M=m} P_{\beta}(\om)\, \D\om,
\ee
where
\be
P_{\beta}(\om)=\frac{e^{-\beta H(\om)}}{Z(\beta)}, 
\label{eqcan1}
\ee
with
\be
Z(\beta)=\int_{\Lambda}e^{-\beta H(\om)}\, \D\om
\ee
as the partition function. The parameterization of the different distributions reflects of course the parameters defining each ensemble: the mean energy $u$ (or the energy $Nu$) for the ME, and the inverse temperature $\beta$ (or temperature $T=(k_B \beta)^{-1}$ with $k_B$ the Boltzmann constant) for the CE.

\section{Mixture of ensembles}

It is obvious from the construction and physical interpretation of the ME and CE that $P^u(m)$ and $P_\beta(m)$ are in general two very different probability distributions of the same macrostate. The point about equivalence of ensembles, however, is that the typical values of $M$ determined by these distributions, which define respectively $\cE^u$ and $\cE_\beta$, may be equivalent in the sense thatm for a given $u$, there might be a given $\beta$ such that $\cE^u=\cE_\beta$. The essential idea involved in establishing this equivalence, which can be traced back to Gibbs \cite{gibbs1902}, is that the CE is a probabilistic mixture of MEs. 

To explain the meaning of this last statement, simply consider the Gibbs canonical distribution of Eq.~(\ref{eqcan1}). Since this distribution depends only on $\beta$ and $H(\om)$, it is clear that all microstates $\om$ having the same energy have the same probabilistic weight, which implies that the conditional probability distribution $P_\beta(\om|h=u)$, obtained by conditioning $P_\beta(\om)$ on the set of microstates $\om$ such that $h(\om)=u$, must be uniform over that constrained set of microstates. Since $P^u(\om)$ has exactly this property, we must therefore have $P_\beta(\om|u)=P^u(\om)$ for all $\om\in\Lambda$ and by extension $P_\beta(m|u)=P^u(m)$. 

With this result, we can use Bayes's Theorem to write
\be
P_\beta(m)=\int_\reals P_\beta(m|u)\, P_\beta(u)\, du=\int_\reals P^u(m)\, P_\beta(u)\, du,
\label{eqpart2}
\ee
where $P_\beta(u)$ is the probability distribution of the mean energy $h$ in the CE. In terms of the density of states $\Omega(u)$, this distribution is simply given by
\be
P_\beta(u)=\frac{\Omega(u)\, e^{-\beta N u}}{Z(\beta)}.
\ee
The second equality obtained in (\ref{eqpart2}) shows that the CE is a superposition of MEs weighted by the canonical probability distribution $P_\beta(u)$ of the mean energy. It is this superposition, which is valid for all $N$, that we refer to as a probabilistic mixture of MEs. 

\section{Results}

The idea that the CE is a superposition of MEs is simple but far reaching because it implies, by taking the thermodynamic limit, that the concentration points of $P_\beta(m)$ are the concentration points of $P^u(m)$ where $P_\beta(u)$ concentrates. In other words, $\cE_\beta$ must be given by all the sets $\cE^u$ such that $u$ is an equilibrium value of the mean energy in the CE at inverse temperature $\beta$. To put this statement in mathematical form, let us denote by $\cU_\beta$ the set of these mean energies realized at equilibrium in the CE at $\beta$. Then we obtain from Eq.~(\ref{eqpart2}):
\be
\cE_\beta=\bigcup_{u\in\cU_\beta}  \cE^u.
\label{eqdec1}
\ee

\begin{figure}[t]
\centering
\resizebox{2.3in}{!}{\includegraphics{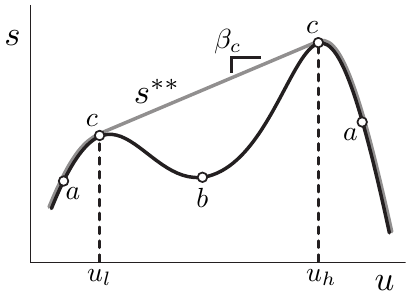}}
\caption{Microcanonical entropy function $s(u)$ (black) and its concave envelope $s^{**}(u)$ (grey). (a) Strictly concave points of $s$; (b) nonconcave points of $s$; (c) Concave points of $s$ that are not strictly concave.}
\label{figcon1}
\end{figure}

This set covering result is a central result of this paper. The equivalence or nonequivalence of the ME and CE follows from this result by determining whether $\cU_\beta$ contains one or more equilibrium mean energies $u$ and whether a given $u$ is a member of $\cU_\beta$ for some $\beta\in\reals$. These questions are known to be answered \cite{touchette2004b} by the concavity properties of the microcanonical thermodynamic entropy or entropy density, defined by the usual limit
\be
s(u)=\lim_{N\ra\infty}\frac{1}{N}\ln \Omega(u).
\ee
We state next the essence of the relationship between the concavity of $s(u)$ and the form of $\cU_\beta$, and then state the consequence of this relationship for the equivalence of the ME and CE as derived from Eq.~(\ref{eqpart2}). For the definition of concave points of $s(u)$, see \cite{touchette2004b}. A basic reference guide to this definition, which is sufficient for the purpose of this paper, is given in Fig.~\ref{figcon1}.

As in \cite{ellis2000}, we are led to consider three cases:

\textbf{Case 1~(Equivalence):} If the entropy $s$ is strictly concave at $u$ (point $a$ in Fig.~\ref{figcon1}), then there exists an inverse temperature $\beta\in\reals$ such that $\cU_\beta$ is a singleton set containing only $u$. For this inverse temperature, Eq.~(\ref{eqpart2}) then reduces to $\cE^u=\cE_\beta$. In this case, the equilibrium states of $M$ in the ME are all realized as equilibrium states in the CE.

\textbf{Case 2~(Nonequivalence):} If $s$ is nonconcave at $u$ (point $b$ in Fig.~\ref{figcon1}), then $u\notin\cU_\beta$ for all $\beta\in\reals$, so that $\cE^u\neq \cE_\beta$ for all $\beta\in\reals$. In fact, it can be proved in this case that $\cE^u\cap \cE_\beta=\emptyset$ for all $\beta\in\reals$. 

The physical interpretation of this case is that the equilibrium states obtained in the ME for values of $u$ where $s(u)$ is nonconcave are not realized as equilibrium states in the CE for any temperature \cite{touchette2004b}. We call these states \emph{microcanonical nonequivalent states}. Models having such states include the mean-field Potts model \cite{costeniuc2005a}, the mean-field Blume-Emery-Griffiths model \cite{ellis2004}, and the point-vertex model of 2D turbulence \cite{ellis2002}.

\textbf{Case 3~(Partial equivalence):} If $s$ is concave but not strictly concave at $u$ (point $c$ in Fig.~\ref{figcon1}), then $u\in\cU_\beta$ for some $\beta\in\reals$, but $u$ is not the only member of $\cU_\beta$, which implies from Eq.~(\ref{eqpart2}) that $\cE^u\subseteq \cE_\beta$.

This last result follows because $\cE_\beta$ in this case is made up of different microcanonical equilibrium states at different mean energies, one of which is $u$. Thus $\cE_\beta$ contains all the equilibrium states of the ME at $u$ but may also contain more equilibrium states not seen in $\cE^u$; hence the name ``partial equivalence''. If any of the many sets that compose $\cE_\beta$ is different from $\cE^u$, which is expected since different energies should give rise to different equilibrium states, then the previous result is strengthened to $\cE^u\subset \cE_\beta$. That is, in this case $\cE_\beta$ does contain equilibrium states that are not contained in $\cE^u$.

\section{Relationship with phase transitions}

Partial equivalence of ensembles is a boundary case between the equivalent and nonequivalent cases, which corresponds physically to a phase coexistence accompanying a first-order phase transition in the CE \cite{touchette2004b}. To understand this point, consider the two points labelled by $c$ in Fig.~\ref{figcon1}. These two points are not strictly concave because they share the same supporting line, which corresponds to the line of the concave envelope $s^{**}(u)$ of $s(u)$. Let us denote the mean energies of these points by $u_l$ and $u_h$, respectively, and let $\beta_c$ be the slope of their supporting line. The non-strict concavity of these points then implies that $P_{\beta_c}(u)$ is a bimodal distribution with two equal-height peaks at $u_l$ and $u_h$ \cite{touchette2004b}, so that $\cU_{\beta_c}=\{u_l,u_h\}$ and $\cE_{\beta_c}=\cE^{u_l}\cup \cE^{u_h}$. This shows that the CE at $\beta_c$ is composed of many equilibrium states of the ME. If we regard each of these microcanonical sets as a ``pure phase'' of the ME, then we can say that the CE shows phase coexistence at $\beta_c$. 

This result is consistent with the fact that the CE shows a first-order phase transition at $\beta_c$. Indeed, it is known that if the entropy $s(u)$ is nonconcave, for example as in Fig.~\ref{figcon1}, then the canonical free energy,
\be
\varphi(\beta)=\lim_{N\ra\infty}-\frac{1}{N}\ln Z(\beta),
\ee
is nondifferentiable at some point, corresponding in Fig.~\ref{figcon1} to $\beta_c$. We thus see that the set of equilibrium states of any system in the CE is, at a point of first-order phase transition in the CE, the combination of two or more sets or ``phases'' of equilibrium states of the ME. This does not imply that each equilibrium state in the CE is itself a superposition or a mixture of two or more equilibrium states of the ME; only that the whole set of equilibrium states in the CE is a such a mixture.

Note, finally, that the connection between nonconcave entropies and nondifferentiable free energies also goes in the opposite direction, in that, if $\varphi(\beta)$ is everywhere differentiable, then $s(u)$ is necessarily strictly concave. We therefore obtain as a general physical result that the ME and CE are equivalent at the macrostate level for any systems that do not have  first-order phase transitions.

\section{Comparison with previous results}

The result of Eq.~(\ref{eqpart2}) and the three cases stated above were first obtained, as mentioned before, by Ellis, Haven and Turkington \cite{ellis2000}. The contribution of the present paper is to re-derive these results under much more general conditions than originally done. All that is required here is for the distributions $P^u(m)$ and $P_\beta(m)$ to exist, so that the equilibrium sets $\cE^u$ and $\cE_\beta$ can be constructed. This is a very weak assumption, which amounts essentially to saying that the results of \cite{ellis2000} actually hold provided that the thermodynamic limit of the ME and CE is well defined. In fact, it can be proved that the existence of $\cE^u$ and $\cE_\beta$ is essentially equivalent to the existence of $s(u)$ and $\varphi(\beta)$, respectively, so that our results about the equivalence and nonequivalence of the ME and CE simply hold provided these thermodynamic functions exist.

In \cite{ellis2000}, the same results were obtained by comparison under two technical conditions. The first is that the mean energy function $h(\om)$ can be rewritten as a function of some macrostate $M(\om)$ in the thermodynamic limit. Mathematically, this amounts to saying that there exists a function $\rh(m)$ of the macrostate $M$, called the energy representation function, such that
\be
\lim_{N\ra\infty} | h(\om)- \rh(M(\om))|=0
\ee
uniformly for all $\om\in\Lambda$. The second condition is that there exists an entropy function $\rs(m)$ associated with $M$. This macrostate entropy is defined similarly as $s(u)$ by replacing the density of states $\Omega(u)$ with the density of states $\Omega(m)$ associated with the microstates having a given macrostate value $M(\om)=m$ (see \cite{ellis2000} for the precise definition).

These two conditions limit greatly the type of systems for which the equivalence of the ME and CE could be analyzed. The first condition in particular is akin to a mean-field requirement, known to be satisfied for mean-field systems and certain long-range interacting systems \cite{ellis2000,ellis2004,barre2005}. It does not apply in a uniform sense to gravitational systems, among other long-range systems, and certainly not for systems with short-range interactions. Moreover, even for mean-field or long-range systems, there is typically only one macrostate that qualifies as a macrostate for which $\rh$ and $\rs$ exist, which means that the equivalence of the ME and CE can only be discussed for this particular macrostate.

In our version of the equivalence and nonequivalence results, these restrictions are completely lifted: we can discuss the equivalence or nonequivalence of the ME and CE for \textit{any} systems and \textit{any} macrostates or observables provided that the equilibrium values of these observables can be defined in each ensemble in the thermodynamic limit. When this is the case, the ME and CE are equivalent essentially if and only if the microcanonical entropy is concave in the thermodynamic limit as a function of the mean energy.

\section{Application}

To illustrate the generality of our results, compared to those of \cite{ellis2000}, we now discuss the equivalence of the ME and CE for the $\alpha$-Ising model, defined by the Hamiltonian
\be
H=\frac{J}{N^{1-\alpha}}\sum_{i>j=1}^N\frac{1-S_iS_j}{|i-j|^\alpha},
\ee
where $S_i=\pm 1$ is a spin variable. This model has been extensively studied because of its rich phase transition behavior \cite{dyson1969,fisher1972,frohlich1982,luijten1997}, which includes a suspected Kosterlitz-Thouless transition for $\alpha=2$ \cite{thouless1969,kosterlitz1976}. For $J>0$ and $0\leq\alpha<1$, Barr\'e \textit{et al}.~\cite{barre2005} were able to show that the model admits an energy representation function $\rh$ and macrostate entropy $\rs$ for a particular macrostate, namely, the macroscopic magnetization profile $m(x)$ (see \cite{barre2005} for the precise definition of this macrostate). Given that the entropy $s(u)$ is concave as a function of $u$ for these parameters, they then concluded that the ME and CE are equivalent at the level of $m(x)$. 

For $\alpha\geq 1$, there is no known energy representation function, so the equivalence of the ME and CE cannot be discussed for this case using the results of \cite{ellis2000}. However, the entropy $s(u)$ exists for these parameters and is everywhere concave, since it is known that the model does not show any first-order phase transitions. From our results, we therefore conclude that the ME and CE are equivalent for this model at the level of $m(x)$. More importantly, they are also equivalent at the level of any macrostate, such as the standard magnetization or the spin distribution, which do not satisfy the requirements of \cite{ellis2000}. Similar conclusions can be reached for other models and macrostates, including short-range models, such as the Ising model at the level of its magnetization, and long-range models such as self-gravitating particle models \cite{note1}. 

\section{Idea of the proof}

We conclude by giving an overview of how the results of this paper are proved at a rigorous level. The first step, following \cite{ellis2000}, is to use large deviation theory \cite{touchette2009} to derive a large deviation principle (LDP) for each of the distributions $P^u(m)$ and $P_\beta(m)$, and to use these LDPs to define (in the thermodynamic limit) the sets $\cE^u$ and $\cE_\beta$ of equilibrium values of the macrostate $M$ in the ME and CE, respectively. 

Heuristically, we express these LDPs by the approximations
\be
P^u(m)\approx e^{-N I^u(m)}
\ee 
and
\be
P_\beta(m)\approx e^{-N I_\beta(m)},
\ee 
which are assumed to hold with sub-exponential correction terms in the limit $N\ra\infty$. Given the rate functions $I^u(m)$ and $I_\beta(m)$, the sets $\cE^u$ and $\cE_\beta$ are defined, respectively, as the sets of global minima of $I^u(m)$ and $I_\beta(m)$ \cite{ellis2000}. 

Next, we use the idea that the CE is a probabilistic mixture of MEs, i.e., Eq.~(\ref{eqpart2}), to obtain the following large deviation approximation:
\be
P_\beta(m)\approx\int_\reals e^{-N \{ I^u(m)+J_\beta(u)\}}\, du,
\label{eqintapp1}
\ee
which involves the rate function $I^u(m)$ of $P^u(m)$ and the rate function $J_\beta(u)$ associated with the LDP of the distribution $P_\beta(u)$ of the mean energy $h$ in the CE:
\be
P_\beta(u)\approx e^{-NJ_\beta(u)}.
\ee
In terms of $s(u)$ and $\varphi(\beta)$, $J_\beta(u)$ is explicitly given by \cite{touchette2009}:
\be
J_\beta(u)=\beta u-s(u)-\varphi(\beta).
\ee 
By heuristically approximating the integral (\ref{eqintapp1}) using Laplace's method, we are then led to the following representation formula connecting the canonical and microcanonical rate functions:
\be
I_\beta(m)=\inf_{u\in\reals}\{I^u(m)+J_\beta(u)\}.
\label{eqrep1}
\ee 
Interestingly, this representation formula connects not only the equilibrium states of the ME and CE, but also the fluctuations of $M$ around these equilibrium states in each ensemble via the fluctuations of $h$ in the CE.

With the formula (\ref{eqrep1}) and the explicit expression of $J_\beta(u)$, we finally arrive at Eq.~(\ref{eqpart2}) by noting that $\cU_\beta$ is the set of global minima of $J_\beta(u)$ and, from there, at the three cases of equivalence mentioned before by studying these global minima in relation to the concavity of $s(u)$. We leave the details of this last step, which involves results from convex analysis, and the rigorous justification of all the approximations mentioned above to a more technical paper \cite{note1}.

\section{Summary}

We have shown in this paper that the equivalence of the ME and CE at the level of their equilibrium states is determined by the concavity of the microcanonical entropy $s(u)$ in a model-independent way. Such a relationship between the thermodynamic and macrostate properties of equilibrium systems was known to hold for mean-field and some long-range systems. Here we have shown that it holds for any many-body system and any observable, so long as their equilibrium behavior is well defined. The same relationship can be derived for other ensembles that are dual in the sense of the ME and CE (e.g., the CE and grand-canonical ensemble or the fixed-magnetization and fixed-magnetic field ensembles) simply by replacing $\cE^u$, $\cE_\beta$ and $s(u)$ with the appropriate sets of equilibrium states and entropy, respectively. The entropy determining the equivalence of the CE and grand-canonical ensemble, for example, is the entropy $s(\rho)$ expressed as a function of the particle density $\rho$. Finally, our results can also be generalized to nonequilibrium ensembles (see, e.g., \cite{touchette2009,evans2005a,jack2010b}), as well as quantum systems when observables that commute with the energy are considered \cite{kastner2010}.

\newpage
\begin{acknowledgments}
I wish to thank R. Nerattini, C. Nardini, G. L. Sewell and R. J. Harris for useful comments. This work was partially supported by an Interdisciplinary Fellowship from RCUK.
\end{acknowledgments}


\end{document}